\documentclass[11pt]{article}
\usepackage{aas_macros}
\usepackage{authblk}
\usepackage{amssymb}
\usepackage{enumitem}
\usepackage{fancyhdr}
\usepackage{geometry}
\usepackage{graphicx}
\usepackage[pdftex,
            pdfauthor={Hincks et al.},
            pdftitle={NDRIO White Paper: Envisioning DRI for the Simons Observatory},
           ]{hyperref}
\usepackage[utf8]{inputenc}
\usepackage{multicol}
\usepackage[authoryear]{natbib}
\usepackage{siunitx}
\usepackage{times}
\usepackage{titlesec}
\usepackage{xcolor}

\definecolor{darkblue}{rgb}{0,0,0.4}
\hypersetup{
  colorlinks,%
  citecolor=darkblue,%
  filecolor=black,%
  linkcolor=darkblue,%
  urlcolor=darkblue
}

\newenvironment{itemize*}%
  {\begin{itemize}%
    \setlength{\itemsep}{0pt}%
    \setlength{\parskip}{0pt}}%
  {\end{itemize}}

\title{NDRIO White Paper: Envisioning Digital Research Infrastructure for the Simons Observatory}
\author[1]{Adam~D.~Hincks\thanks{Corresponding author. Email: adam.hincks@utoronto.ca}}
\author[2]{Simone~Aiola\thanks{Simons Observatory Data and Pipeline Project Leader}}
\author[3]{J.~Richard~Bond}
\author[4]{Erminia~Calabrese\thanks{Simons Observatory Theory and Analysis Committee Chair}}
\author[5]{Andrei~Frolov}
\author[3]{Jos\'e~Tom\'as~G\'alvez~Ghersi}
\author[1,6]{Ren\'ee~Hlo\v{z}ek}
\author[7,8]{Matthew~Johnson}
\author[8]{Mathew~S.~Madhavacheril}
\author[8]{Moritz~M\"{u}nchmeyer}
\author[9]{Lyman~A.~Page}
\author[10]{Jonathan~Sievers}
\author[9]{Suzanne~T.~Staggs\thanks{Simons Observatory Spokesperson}}
\author[11]{Alexander~Van~Engelen}
\affil[1]{Department of Astronomy \& Astrophysics, University of Toronto, 50 St George St, Toronto ON M5S 3H4, Canada}
\affil[2]{Center for Computational Astrophysics, Flatiron Institute, 162 5th Avenue, New York, NY 10010 USA}
\affil[3]{Canadian Institute for Theoretical Astrophysics, University of Toronto, 60 St. George Street, Toronto ON M5S 3H8, Canada}
\affil[4]{School of Physics and Astronomy, Cardiff University, The Parade, Cardiff CF24 3AA, UK}
\affil[5]{Physics Department, Simon Fraser University, 8888 University Drive, Burnaby BC V5A 1S6, Canada}
\affil[6]{Dunlap Institute, University of Toronto, 50 St George Street, Toronto ON M5S 3H4, Canada}
\affil[7]{Department of Physics, York University, 4700 Keele Street, Toronto ON M3J 1P3, Canada}
\affil[8]{Perimeter Institute for Theoretical Physics, 31 Caroline Street N, Waterloo ON N2L 2Y5, Canada}
\affil[9]{Joseph Henry Laboratories of Physics, Jadwin Hall, Princeton University, Princeton, NJ
08544, USA}
\affil[10]{Physics Department, McGill University, 3600 Rue University,  Montr\'{e}al QC H3A 2T8, Canada}
\affil[11]{School of Earth and Space Exploration, Arizona State University, Tempe, AZ 85287, USA}
\date{December 15, 2020}
\begin{document}

\maketitle

\begin{abstract}
\noindent Observations of the cosmic microwave background (CMB) are an incredibly fertile source of information for studying the origins and evolution of the Universe. Canadian digital research infrastructure (DRI) has played a key role in reducing ever-larger quantities of raw data into maps of the CMB suitable for scientific analysis, as exemplified by the many scientific results produced by the Atacama Cosmology Telescope (ACT) over the past decade. The Simons Observatory (SO), due to start observing in 2023, will be able to measure the CMB with about an order of magnitude more sensitivity than ACT and other current telescopes. In this White Paper we outline how Canadian DRI under the New Digital Research Infrastructure Organization (NDRIO) could build upon the legacy of ACT and play a pivotal role in processing SO data, helping to produce data products that will be central to the cosmology community for years to come. We include estimates of DRI resources required for this work to indicate what kind of advanced research computing (ARC) would best support an SO-like project. Finally, we comment on how ARC allocations could be structured for large collaborations like SO and propose a research data management (RDM) system that makes public data releases available not only for download but also for direct analysis on Canadian DRI.
\end{abstract}

\section{Introduction}

For more than a decade, Canadian digital research infrastructure (DRI) has been central to advances in cosmology through the study of the cosmic microwave background (CMB), the oldest light in the Universe. Compute Canada (CC) has supported the analysis and processing of microwave data from experiments such as the \textit{Planck} satellite, the South Pole Telescope (SPT), the SPIDER B-mode experiment, the Atacama B-Mode Search (ABS) and the Atacama Cosmology Telescope (ACT). The challenges of analysing data sets from these diverse experiments differ. As an example, we will take as a baseline the computational needs of ACT, which has been a significant user of Canadian DRI since 2009, and then discuss the transition to its next-generation successor, the Simons Observatory (SO). Future telescopes such as the Cerro Chajnantor Atacama Telescope-prime (CCAT-prime), the \textit{LiteBIRD} satellite and the CMB Stage Four (CMB-S4) telescopes \citep[c.f.,][]{hlozek/etal:2019} will have analogous computing needs, though with varying data volumes. The recent Canadian Astronomical Society Long Range Plan for the coming decade listed Canadian participation in a large ground-based CMB survey like CMB-S4 or SO, as well as the space-based LiteBIRD, as priorities \citep{barmby/etal:2020}. DRI support will be critical to realising these goals.

\begin{figure}[tb]
    \centering
    \includegraphics{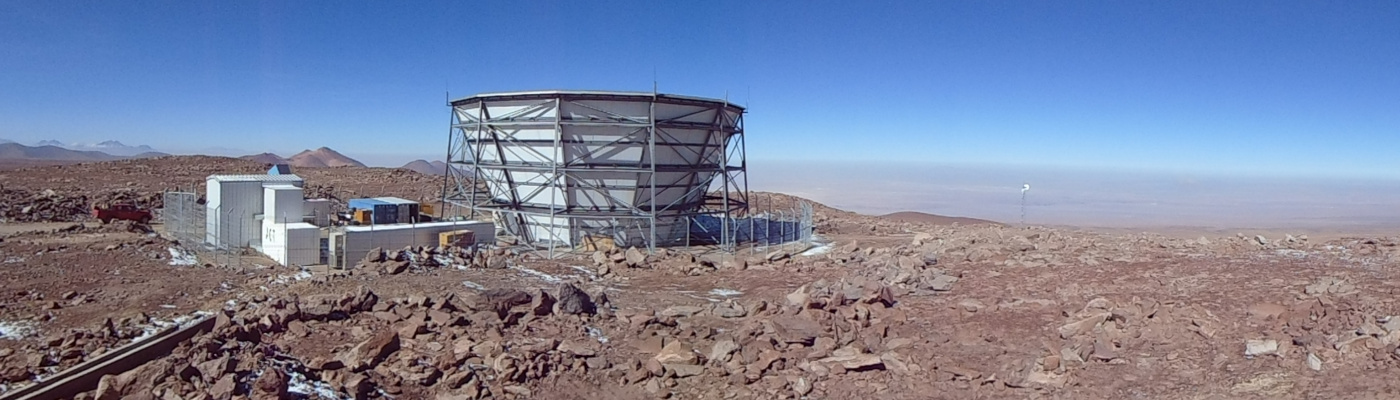}
    \caption{The Atacama Cosmology Telescope. The Simons Observatory will be at the same site. (Photo: A. Hincks.)}
    \label{fig:act}
\end{figure}

\begin{figure}[tb]
    \centering
    \includegraphics[width=5in]{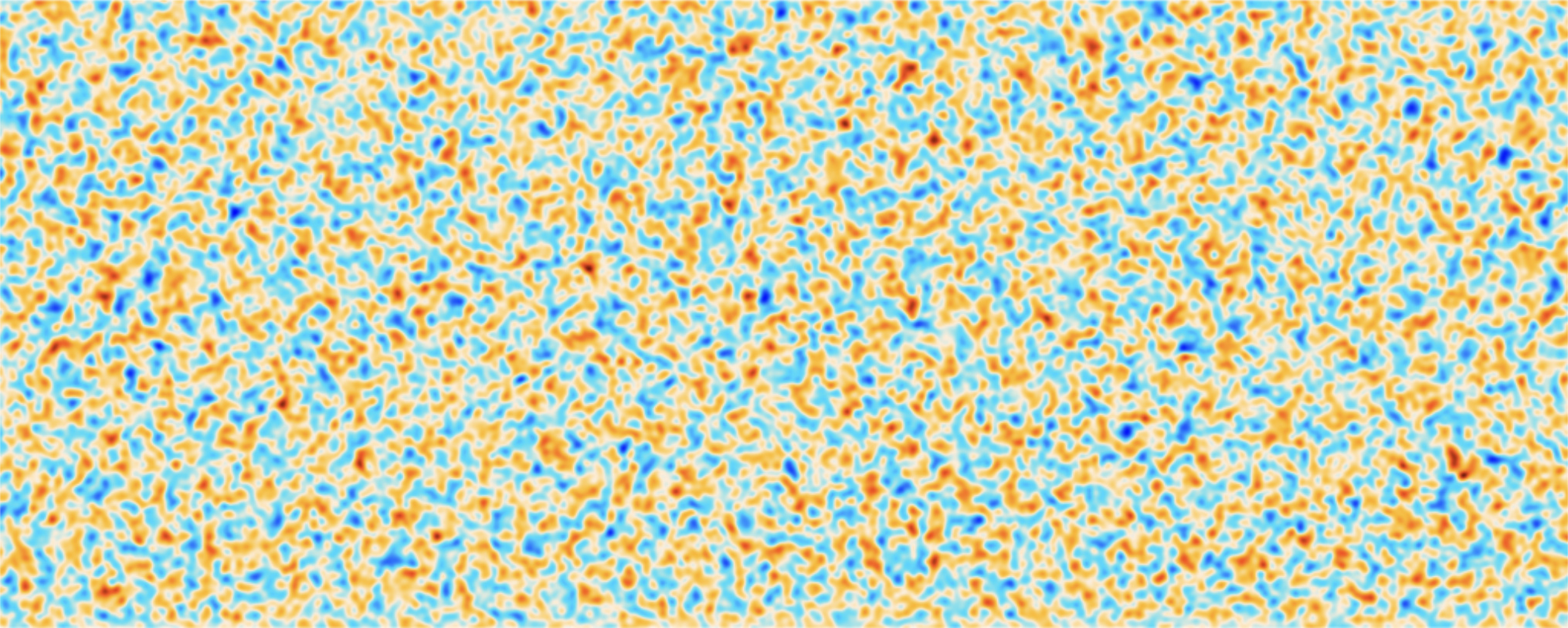}
    \caption{A section of the map of E-mode polarisation of the CMB from ACT's DR4 \citep{aiola/etal:2020,choi/etal:2020}. This map was made with SciNet's Niagara cluster. The width shown here is 25\,degrees, corresponding to 20 billion light years. The colours represent small variations in the amount of polarisation ($\pm25\,\si{\micro\kelvin}$) and the size of these features was used to make a new measurement of the Hubble constant. (Image credit: ACT Collaboration.)}
    \label{fig:dr4_cmb_image}
\end{figure}

ACT is a 6\,m telescope in northern Chile that observes the CMB at high resolution over wide areas in multiple frequencies \citep[Fig.~\ref{fig:act};][]{swetz/etal:2011,thornton/etal:2016,henderson/etal:2016}. Raw data have been reduced to maps of the sky using resources managed by Toronto's SciNet Consortium, starting with the GPC cluster in 2009 and continuing on its successor Niagara. These maps have enabled almost all of the $\sim$70 scientific papers produced by the collaboration, including groundbreaking results such as the the first detection of the power spectrum of gravitational lensing of the CMB \citep{das/etal:2011,sherwin/etal:2011} and the first measurement of the bulk motion of massive clusters of galaxies using the kinematic Sunyaev-Zeldovich effect \citep{hand/etal:2012}. The latest public data release from the collaboration in 2020, Data Release 4 (DR4; see Fig.~\ref{fig:dr4_cmb_image}),\footnote{Available at \url{https://lambda.gsfc.nasa.gov/product/act/actpol_prod_table.cfm}.} has generated great interest in the cosmology research community and received widespread coverage in the international press \citep[e.g.,][]{mortillaro:2020,amos:2020}. The maps and statistical data products released as part of DR4 yielded excellent agreement with the standard model of cosmology and provided the first independent check by a high resolution experiment of the Hubble constant with CMB data \citep{aiola/etal:2020,choi/etal:2020}. Numerous ACT papers based on DR4 and preliminary versions of the next data release, DR5, have appeared or are in preparation.\footnote{See \url{https://act.princeton.edu/publications}.}

ACT and other CMB experiments have not only benefited from Canadian DRI but have spurred its development: one of the reasons SciNet was created was to provide the computational resources required by these collaborations. Today, a new opportunity for DRI innovation in Canada is presented by the Simons Observatory (SO), an upcoming CMB experiment. For its initial five-year survey (2024--29), SO will consist of one 6\,m large aperture telescope (LAT) and three 50\,cm small aperture telescopes \citep[SAT;][]{lee/etal:2019}. Together they will deliver an order-of-magnitude better sensitivity than current CMB experiments, with the lower-resolution SATs geared towards detecting the signature of primordial gravity waves that would be decisive evidence that the Universe underwent a rapid period of inflation in the first instants after the big bang, while the higher-resolution LAT will measure the total mass of the three neutrino species, learn about the nature of dark matter and dark energy, discover many thousands of clusters of galaxies, and much more \citep{ade/etal:2019}. The increase in sensitivity provided by SO is enabled by fielding more than ten times the number of detectors in ACT, but this advance comes with a commensurate increase in the volume and complexity of the data that need to be processed.

We propose that the New Digital Research Infrastructure Organization (NDRIO) help enable SO science by providing support to make CMB maps from the LAT data. This would build on Canada's world-class leadership in developing CMB experiments and processing and analysing their data. The DRI required by SO represents a significant increase over current needs, and would be an impressive contribution to the collaboration whose data products will be essential to cosmology research for the coming decade. After a review of the current use of Canadian DRI for ACT data processing and the particular challenges we are beginning to face (\S\ref{sec:today}), we present the DRI required for the reduction of SO LAT data (\S\ref{sec:future}). Finally, we make concrete recommendations for how advanced research computing (ARC) and research data management (RDA) can be integrated to meet the needs of a large-scale project like SO (\S\ref{sec:recommendations}).

\section{Current Use of DRI and Associated Challenges}
\label{sec:today}

\paragraph{DRI Resources Currently Required:} In order to contextualise the needs of SO, we provide here a summary of ACT's anticipated use of resources on SciNet's Niagara cluster for the coming year. By far the majority of cycles consist in reducing raw data to maps of the CMB sky in temperature and polarisation (see Table~\ref{tab:current_use}). ACT's $\sim$3,000 detectors produce about 15 terasamples per year, for a total of 130\,TB for the 2017–20 dataset being processed. The map-making software, largely written by ACT researchers, uses efficient parallel algorithms, particularly preconditioned conjugant gradient (PCG) solvers, to complete the enormous matrix operations required to reconstruct the maps while optimally removing the noise from the celestial signal. The non-trivial challenges associated with doing this over a large fraction of the sky (40\%) at $\sim$arcminute resolution have made heavy use of the highly reliable MPI communication provided by Niagara. Map-making of ACT's 2017--20 observations will take about 1280 core years,\footnote{Here and throughout we refer to CPU, rather than GPU, processing.} with about 20\,TB of scratch space needed for storing intermediate data products during the process; the final products are $\sim$5\,TB. Memory needs are relatively modest and Niagara's $\sim$4\,GB RAM per core are quite sufficient. In addition to map-making, Niagara is used to determine the telescope pointing solution using time-series analyses of bright point sources ($\sim$120 core years). Furthermore, interpretation of the final products, especially the gravitational lensing of the CMB, relies on processing of simulated data, requiring about $\sim$250 core-years. To date, all of the foregoing work has been done by about 30 researchers under a CC allocation led by principal investigator (PI) J.~R.~Bond. These researchers, who range from students to senior faculty, are based at both Canadian and international institutions. The same allocation, currently being used near capacity, supports other important CMB projects. The SPIDER collaboration ($\sim$20 researchers) uses it to do similar map-making to ACT on larger angular scales. In the coming year they aim to finalise maps from a 16-day flight of their balloon-borne telescope in 2015 \citep{gualtieri/etal:2018}; together with generating the requisite simulations, this will require about 180 core-years with $\sim$13\,TB of storage space. Finally, about 15 researchers employ the allocation for creating sophisticated models of secondary components in the CMB, such as clusters of galaxies, point sources and the effects of possible non-Gaussian physics in the early Universe. These models inform not only analysis of the CMB but also other cosmological datasets, such as those from the Canadian Hydrogen Intensity Mapping Experiment \citep{bandura/etal:2014}  or the \textit{Euclid} satellite \citep{racca/etal:2016}, which have significant Canadian membership. In the coming year this work will take about 250 core years and 12\,TB of storage.

\begin{table}[t]
    \centering
    \begin{tabular}{|l|c|c|c|c|c|c|}
      \hline
        \textbf{Project} & \multicolumn{3}{c|}{\textbf{ACT}} & \textbf{SPIDER} & \textbf{Theory Sims} & \textbf{Total} \\
        & \textit{map-making} & \textit{pointing} & \textit{lensing analysis} & & & \\
        \hline
        \textbf{Core Years} & 1280 & 120 & 250 & 180 & 250 & 2080 \\
        \hline
        \textbf{Storage (TB)} & \multicolumn{3}{c|}{130} & 13 & 12 & 155 \\
        \hline
        \textbf{No. Researchers} & \multicolumn{3}{c|}{$\sim$30} & $\sim$20 & $\sim$15 & $\sim$65 \\
        \hline
    \end{tabular}
    \caption{Anticipated use of DRI resources for CMB research on SciNet's Niagara cluster in 2021--22.}
    \label{tab:current_use}
\end{table}

\paragraph{Current Limitations and Challenges:} Thus far, the CC allocation model has served the CMB community extremely well. However, two limitations of the current system are beginning to emerge. First, CC allocations are tied to single PIs, which can be unwieldy for large collaborations---or, in the scenario described above, for more than one large collaboration sharing an allocation. The allocation PI is not necessarily the PI of the collaboration; if, for some reason, the former must step down, it is not clear how the collaboration's work can move forward without the significant interruption of having a new PI apply for a new allocation. Furthermore, within just a single collaboration like ACT there are several distinct computational tasks (c.f., Table~\ref{tab:current_use}) that are nonetheless tightly coupled. Currently there are two possibilities: keep them under a single allocation, where they may compete for cycles, or put them under separate allocations, where it becomes more difficult to integrate their results. Neither is fully satisfactory. Second, the one-year allocation length is short compared with the years of continual effort now required to process a single CMB dataset. Even though allocation renewals can be `fast tracked' for two more years after an initial successful application, the cadence of allocation cycles is becoming incommensurate with experiment timelines. To address these limitations, we propose the introduction of a `team allocation', described in \S\ref{sec:recommendations}, below.

\section{Looking to the Future: Computing Requirements for the SO LAT}
\label{sec:future}

\paragraph{Building on the Legacy of ACT:} The SO LAT is similar to ACT in terms of sky coverage ($\sim$40\%), angular resolution and frequencies observed. Doing SO map-making on Canadian DRI thus makes eminent practical sense, since much of the research software (RS) will be similar and build on the demonstrated success of ACT computation. The LAT will do a five years survey using about ten times more detectors than ACT, and this increased data volume will be the main challenge for DRI. It requires a major step forward in terms of storage, computing cycles and total amount of resources used per single job. Let us now quantify these requirements.

\paragraph{Required DRI for the SO LAT:} The SO LAT will produce roughly 1.5\,PB of raw time-ordered data over its five year survey.\footnote{This figure is less than one gets from scaling the ACT volume by a 10$\times$ increase in detectors, due to lower sampling rates on SO.} Each year of data (called a `season') will be continuously analysed for at least two subsequent years; indeed it is conceivable that all 1.5\,PB will need to remain `spinning'---or if on tape, seamlessly transferred to disk upon read---for a seven year period during which the full dataset will be reprocessed with improvements to the reduction pipelines.\footnote{The number seven comes from the full period of official SO operations: five years of observations and data reduction, and two extra years to complete the final delivery of the products.} We base this upon experience from ACT that as sensitivity grows with increased data volume, new systematic effects emerge that need to be understood, modelled and accounted for with sophisticated upgrades to the software pipelines. In the high-precision realm of SO science, this process will be essential. Final science-quality products and associated simulations will also require storage at the level of a few hundred TB and will need to be made accessible to a large number of collaborators and to the broader community. In addition to the large amount of storage, a significant increase in computational cycles is required. From estimates based on the performance of the ACT code, we expect a computational cost of 300 core years to reduce a single season of LAT data.  Additional RAM (on the order of 10's of TB total) can reduce this somewhat, both by eliminating repeated calculations and enabling more algorithms. Nonetheless, the CPU usage number only represents the final pass at the data: before production runs, there is a critical phase of data exploration and characterisation of systematics, as mentioned above. Even though only a subset of the yearly data is processed, it must be done multiple times. Our rough estimate is that such exploratory work amounts to about 20 times the cycles of a full season run. This brings the total to about 6,000 core-years per season of data. In the scenario of a cumulative analysis of the dataset, the annual requirement will have increased to roughly 30,000 core-years by 2029, or 18 times higher than the anticipated cycles needed for ACT processing in 2021--22 (see Table~\ref{tab:current_use}).  

\paragraph{Coordination with International DRI:} SO is a large international collaboration, some of whose members have access to non-Canadian DRI, or are in positions to make strong bids in competitive applications for their use. Based on previous experience, the best approach is not to make use of a single resource but to strategically coordinate among them: for instance, while almost all the principal map-making for ACT has been done on Niagara, other computing has been done on clusters at the U.S. National Energy Research Scientific Computing Center (NERSC), at the university of KwaZulu-Natal and at Princeton University. This multi-pronged approach has proven extremely effective for avoiding single-point failures and for distributing diverse types of computation among different DRI. For SO, we envision a scenario where it is not Canadian \textit{or} international DRI, but Canadian DRI \textit{with} international DRI. As a concrete example, we might envision running time-domain simulations of our raw data and map-making of the lower resolution SAT data at NERSC, while using Canadian DRI for map-making of the LAT data as proposed above. The LAT and SATs are sufficiently decoupled that their data can be processed almost completely independently of each other, making such a division of labour trivial. Nonetheless, we should add that the coordination of DRI outlined in this White Paper (i.e., LAT map-making in Canada and other processing at NERSC) is not the only possibility. There are other scenarios in which SAT map-making and other computational tasks are done with NDRIO DRI that would utilise similar resources to those we outline in this White Paper. Our intent has been to focus on a well-motivated plan that provides a clear indication of DRI requirements for upcoming CMB processing.

\section{Bridging the Gap: Recommendations for NDRIO}
\label{sec:recommendations}

Let us now make some specific recommendations for NDRIO that would enable us to capitalise on the singular opportunity to process SO LAT data with Canadian DRI. An obvious \textit{sine qua non} would be securing an allocation on ARC that meets the requirements for storage and computing power outlined above (\S\ref{sec:future}), namely:
\begin{itemize*}
  \item capacity for 30,000 core years of processing (baselined to Niagara's CPUs),
  \item 1.5\,PB of spinning storage during processing,
  \item 1.5\,PB of cold storage for data recovery and
  \item several 100\,TB of storage for final, publicly available data products (maps and associated simulations)
\end{itemize*}

\noindent Additionally, we propose the following features for NDRIO ARC and RDM:

\paragraph{1. Team Allocations:} To overcome the limitation of the single-PI allocation model described in \S\ref{sec:today}, we recommend the addition of a \textit{team allocation} with a 7 year duration. The allocation would be targeted at SO-sized collaborations, and would be applied for and managed by a small group of Canada-based PIs, or team leaders, from within the collaboration. A team allocation would be used for the largest and longest-term computing tasks of the collaboration---in the case of SO, this would consist of map-making. Single PI allocations would still exist, and members of the collaboration would be encouraged to secure their own allocations for short-term, ancillary projects requiring HPC. The length of single-PI allocations need not be as long as team allocations.

\paragraph{2. Collocation of Public Data Hosting and ARC:} In the hybrid allocation model described above, there would have to be RDM capabilities such that members of the collaboration with single-PI allocations could seamlessly access, add to or update data products produced with the team allocation. Most basically this could simply occur through intelligent management of file system permissions, though one might imagine RDM solutions that aid sharing of various stages of data products within a team allocation and across authorised single-PI allocations. However, we propose a much more ambitious vision of RDM that enables collaborations to release final, public data products to the broader research community. Given the enormous size of these products (hundreds of TB for SO: see \S\ref{sec:future}), this would mean more than just hosting files for download as per the current model; though this would still be important, it could be impracticable when the whole dataset is involved. Ideally, anyone with a single-PI or team allocation on NDRIO ARC could directly access these products---in other words, they would be collocated with the ARC. One could even envision RDM that automatically mirrors public data products at multiple HPC facilities across the country for the broadest possible access; furthermore, we recommend that there be a pathway for non-Canadian researchers to apply for NDRIO ARC resources in order to amplify the international impact of the data products by maximising their use.

\section{Conclusions}

For years now, cosmologists have turned to Canadian DRI for the critical work of processing raw time streams from CMB experiments into the science-grade data products that have advanced our knowledge of the Universe. With the advent of next-generation CMB experiments in the coming years, we have an exciting opportunity to build upon this legacy and adapt to the new computational demands that larger observatories present. To illustrate this we have focused on the transition from ACT to the SO LAT, and presented a scenario in which a significant contribution to upcoming CMB science can be achieved with future Canadian DRI within an international context. As mentioned in the Introduction, we believe the recommendations outlined in this paper would benefit not only the SO collaboration, but other future CMB experiments, and provide data products that will enhance our understanding of cosmology for many years to come.

\titleformat*{\paragraph}{\itshape}
\paragraph{Acknowledgements} We thank Michael Nolta for helpful advice and feedback during preparation of this paper.

\setlength{\bibsep}{0pt plus 0.3ex}
\begin{multicols}{2}
\footnotesize{
\bibliography{main}

\begin{thebibliography}{}
\expandafter\ifx\csname natexlab\endcsname\relax\def\natexlab#1{#1}\fi
\providecommand{\url}[1]{\href{#1}{#1}}
\providecommand{\dodoi}[1]{doi:~\href{http://doi.org/#1}{\nolinkurl{#1}}}
\providecommand{\doeprint}[1]{\href{http://ascl.net/#1}{\nolinkurl{http://ascl.net/#1}}}
\providecommand{\doarXiv}[1]{\href{https://arxiv.org/abs/#1}{\nolinkurl{https://arxiv.org/abs/#1}}}

\bibitem[{{Ade} {et~al.}(2019){Ade}, {Aguirre}, {Ahmed}, {Aiola}, {Ali},
  {Alonso}, {Alvarez}, {Arnold}, {Ashton}, {Austermann}, {Awan}, {Baccigalupi},
  {Baildon}, {Barron}, {Battaglia}, {Battye}, {Baxter}, {Bazarko}, {Beall},
  {Bean}, {Beck}, {Beckman}, {Beringue}, {Bianchini}, {Boada}, {Boettger},
  {Bond}, {Borrill}, {Brown}, {Bruno}, {Bryan}, {Calabrese}, {Calafut},
  {Calisse}, {Carron}, {Challinor}, {Chesmore}, {Chinone}, {Chluba}, {Cho},
  {Choi}, {Coppi}, {Cothard}, {Coughlin}, {Crichton}, {Crowley}, {Crowley},
  {Cukierman}, {D'Ewart}, {D{\"u}nner}, {de Haan}, {Devlin}, {Dicker},
  {Didier}, {Dobbs}, {Dober}, {Duell}, {Duff}, {Duivenvoorden}, {Dunkley},
  {Dusatko}, {Errard}, {Fabbian}, {Feeney}, {Ferraro}, {Flux{\`a}}, {Freese},
  {Frisch}, {Frolov}, {Fuller}, {Fuzia}, {Galitzki}, {Gallardo}, {Tomas Galvez
  Ghersi}, {Gao}, {Gawiser}, {Gerbino}, {Gluscevic}, {Goeckner-Wald}, {Golec},
  {Gordon}, {Gralla}, {Green}, {Grigorian}, {Groh}, {Groppi}, {Guan},
  {Gudmundsson}, {Han}, {Hargrave}, {Hasegawa}, {Hasselfield}, {Hattori},
  {Haynes}, {Hazumi}, {He}, {Healy}, {Henderson}, {Hervias-Caimapo}, {Hill},
  {Hill}, {Hilton}, {Hilton}, {Hincks}, {Hinshaw}, {Hlo{\v{z}}ek}, {Ho}, {Ho},
  {Howe}, {Huang}, {Hubmayr}, {Huffenberger}, {Hughes}, {Ijjas}, {Ikape},
  {Irwin}, {Jaffe}, {Jain}, {Jeong}, {Kaneko}, {Karpel}, {Katayama}, {Keating},
  {Kernasovskiy}, {Keskitalo}, {Kisner}, {Kiuchi}, {Klein}, {Knowles},
  {Koopman}, {Kosowsky}, {Krachmalnicoff}, {Kuenstner}, {Kuo}, {Kusaka},
  {Lashner}, {Lee}, {Lee}, {Leon}, {Leung}, {Lewis}, {Li}, {Li}, {Limon},
  {Linder}, {Lopez-Caraballo}, {Louis}, {Lowry}, {Lungu}, {Madhavacheril},
  {Mak}, {Maldonado}, {Mani}, {Mates}, {Matsuda}, {Maurin}, {Mauskopf}, {May},
  {McCallum}, {McKenney}, {McMahon}, {Meerburg}, {Meyers}, {Miller},
  {Mirmelstein}, {Moodley}, {Munchmeyer}, {Munson}, {Naess}, {Nati},
  {Navaroli}, {Newburgh}, {Nguyen}, {Niemack}, {Nishino}, {Orlowski-Scherer},
  {Page}, {Partridge}, {Peloton}, {Perrotta}, {Piccirillo}, {Pisano},
  {Poletti}, {Puddu}, {Puglisi}, {Raum}, {Reichardt}, {Remazeilles},
  {Rephaeli}, {Riechers}, {Rojas}, {Roy}, {Sadeh}, {Sakurai}, {Salatino},
  {Sathyanarayana Rao}, {Schaan}, {Schmittfull}, {Sehgal}, {Seibert}, {Seljak},
  {Sherwin}, {Shimon}, {Sierra}, {Sievers}, {Sikhosana}, {Silva-Feaver},
  {Simon}, {Sinclair}, {Siritanasak}, {Smith}, {Smith}, {Spergel}, {Staggs},
  {Stein}, {Stevens}, {Stompor}, {Suzuki}, {Tajima}, {Takakura}, {Teply},
  {Thomas}, {Thorne}, {Thornton}, {Trac}, {Tsai}, {Tucker}, {Ullom},
  {Vagnozzi}, {van Engelen}, {Van Lanen}, {Van Winkle}, {Vavagiakis},
  {Verg{\`e}s}, {Vissers}, {Wagoner}, {Walker}, {Ward}, {Westbrook},
  {Whitehorn}, {Williams}, {Williams}, {Wollack}, {Xu}, {Yu}, {Yu}, {Zago},
  {Zhang}, {Zhu}, \& {Simons Observatory Collaboration}}]{ade/etal:2019}
{Ade}, P., {Aguirre}, J., {Ahmed}, Z., {et~al.} 2019, \jcap, 2019, 056,
  \dodoi{10.1088/1475-7516/2019/02/056}

\bibitem[{{Aiola} {et~al.}(2020){Aiola}, {Calabrese}, {Maurin}, {Naess},
  {Schmitt}, {Abitbol}, {Addison}, {Ade}, {Alonso}, {Amiri}, {Amodeo},
  {Angile}, {Austermann}, {Baildon}, {Battaglia}, {Beall}, {Bean}, {Becker},
  {Bond}, {Bruno}, {Calafut}, {Campusano}, {Carrero}, {Chesmore}, {Cho.},
  {Choi}, {Clark}, {Cothard}, {Crichton}, {Crowley}, {Darwish}, {Datta},
  {Denison}, {Devlin}, {Duell}, {Duff}, {Duivenvoorden}, {Dunkley},
  {D{\"u}nner}, {Essinger-Hileman}, {Fankhanel}, {Ferraro}, {Fox}, {Fuzia},
  {Gallardo}, {Gluscevic}, {Golec}, {Grace}, {Gralla}, {Guan}, {Hall},
  {Halpern}, {Han}, {Hargrave}, {Hasselfield}, {Helton}, {Henderson},
  {Hensley}, {Hill}, {Hilton}, {Hincks}, {Ho}, {Hubmayr}, {Huffenberger},
  {Hughes}, {Infante}, {Irwin}, {Klein}, {Knowles}, {Kosowsky}, {Lakey}, {Li},
  {Li}, {Lokken}, {Louis}, {Lungu}, {Madhavacheril}, {Maldonado},
  {Mallaby-Kay}, {Marsden}, {McMahon}, {Menanteau}, {Morton}, {Namikawa},
  {Newburgh}, {Nibarger}, {Niemack}, {Nolta}, {Orlowski-Sherer}, {Page},
  {Pappas}, {Partridge}, {Phakathi}, {Prince}, {Puddu}, {Qu}, {Robertson},
  {Rojas}, {Schaan}, {Schillaci}, {Sehgal}, {Sherwin}, {Sierra}, {Sievers},
  {Sikhosana}, {Simon}, {Staggs}, {Stevens}, {Storer}, {Sunder}, {Switzer},
  {Thorne}, {Thornton}, {Trac}, {Treu}, {Tucker}, {Vale}, {Van Engelen}, {Van
  Lanen}, {Vavagiakis}, {Wagoner}, {Wang}, {Wollack}, {Xu}, \&
  {Zhu}}]{aiola/etal:2020}
{Aiola}, S., {Calabrese}, E., {Maurin}, L., {et~al.} 2020, arXiv e-prints,
  arXiv:2007.07288.
\newblock \doarXiv{2007.07288}

\bibitem[{{Amos}(2020)}]{amos:2020}
{Amos}, J. 2020, BBC News.
\newblock \url{https://www.bbc.com/news/science-environment-53420433}

\bibitem[{{Bandura} {et~al.}(2014){Bandura}, {Addison}, {Amiri}, {Bond},
  {Campbell-Wilson}, {Connor}, {Cliche}, {Davis}, {Deng}, {Denman}, {Dobbs},
  {Fandino}, {Gibbs}, {Gilbert}, {Halpern}, {Hanna}, {Hincks}, {Hinshaw},
  {H{\"o}fer}, {Klages}, {Landecker}, {Masui}, {Mena Parra}, {Newburgh}, {Pen},
  {Peterson}, {Recnik}, {Shaw}, {Sigurdson}, {Sitwell}, {Smecher}, {Smegal},
  {Vanderlinde}, \& {Wiebe}}]{bandura/etal:2014}
{Bandura}, K., {Addison}, G.~E., {Amiri}, M., {et~al.} 2014, in Society of
  Photo-Optical Instrumentation Engineers (SPIE) Conference Series, Vol. 9145,
  Ground-based and Airborne Telescopes V, ed. L.~M. {Stepp}, R.~{Gilmozzi}, \&
  H.~J. {Hall}, 914522, \dodoi{10.1117/12.2054950}

\bibitem[{{Barmby} {et~al.}(2020){Barmby}, {Gaensler}, {Dobbs}, {Heyl},
  {Ivanova}, {Lafreni\`{e}re}, {Matthews}, \& {Shapley}}]{barmby/etal:2020}
{Barmby}, P., {Gaensler}, B., {Dobbs}, M., {et~al.} 2020, {Final report of the
  Canadian Astronomical Society’s 2020 Long Range Plan for Canadian Astronomy
  (LRP2020)}, Tech. rep., {Canadian Astronomical Society}.
\newblock
  \url{https://casca.ca/wp-content/uploads/2020/12/LRP2020_December2020-1.pdf}

\bibitem[{{Choi} {et~al.}(2020){Choi}, {Hasselfield}, {Ho}, {Koopman}, {Lungu},
  {Abitbol}, {Addison}, {Ade}, {Aiola}, {Alonso}, {Amiri}, {Amodeo}, {Angile},
  {Austermann}, {Baildon}, {Battaglia}, {Beall}, {Bean}, {Becker}, {Bond},
  {Bruno}, {Calabrese}, {Calafut}, {Campusano}, {Carrero}, {Chesmore}, {Cho},
  {Clark}, {Cothard}, {Crichton}, {Crowley}, {Darwish}, {Datta}, {Denison},
  {Devlin}, {Duell}, {Duff}, {Duivenvoorden}, {Dunkley}, {D{\"u}nner},
  {Essinger-Hileman}, {Fankhanel}, {Ferraro}, {Fox}, {Fuzia}, {Gallardo},
  {Gluscevic}, {Golec}, {Grace}, {Gralla}, {Guan}, {Hall}, {Halpern}, {Han},
  {Hargrave}, {Henderson}, {Hensley}, {Hill}, {Hilton}, {Hilton}, {Hincks},
  {Hlo{\v{z}}ek}, {Hubmayr}, {Huffenberger}, {Hughes}, {Infante}, {Irwin},
  {Jackson}, {Klein}, {Knowles}, {Kosowsky}, {Lakey}, {Li}, {Li}, {Li},
  {Lokken}, {Louis}, {MacInnis}, {Madhavacheril}, {Maldonado}, {Mallaby-Kay},
  {Marsden}, {Maurin}, {McMahon}, {Menanteau}, {Moodley}, {Morton}, {Naess},
  {Namikawa}, {Nati}, {Newburgh}, {Nibarger}, {Nicola}, {Niemack}, {Nolta},
  {Orlowski-Sherer}, {Page}, {Pappas}, {Partridge}, {Phakathi}, {Prince},
  {Puddu}, {Qu}, {Rivera}, {Robertson}, {Rojas}, {Salatino}, {Schaan},
  {Schillaci}, {Schmitt}, {Sehgal}, {Sherwin}, {Sierra}, {Sievers}, {Sifon},
  {Sikhosana}, {Simon}, {Spergel}, {Staggs}, {Stevens}, {Storer}, {Sunder},
  {Switzer}, {Thorne}, {Thornton}, {Trac}, {Treu}, {Tucker}, {Vale}, {Van
  Engelen}, {Van Lanen}, {Vavagiakis}, {Wagoner}, {Wang}, {Ward}, {Wollack},
  {Xu}, {Zago}, \& {Zhu}}]{choi/etal:2020}
{Choi}, S.~K., {Hasselfield}, M., {Ho}, S.-P.~P., {et~al.} 2020, arXiv
  e-prints, arXiv:2007.07289.
\newblock \doarXiv{2007.07289}

\bibitem[{{Das} {et~al.}(2011){Das}, {Sherwin}, {Aguirre}, {Appel}, {Bond},
  {Carvalho}, {Devlin}, {Dunkley}, {D{\"u}nner}, {Essinger-Hileman}, {Fowler},
  {Hajian}, {Halpern}, {Hasselfield}, {Hincks}, {Hlozek}, {Huffenberger},
  {Hughes}, {Irwin}, {Klein}, {Kosowsky}, {Lupton}, {Marriage}, {Marsden},
  {Menanteau}, {Moodley}, {Niemack}, {Nolta}, {Page}, {Parker}, {Reese},
  {Schmitt}, {Sehgal}, {Sievers}, {Spergel}, {Staggs}, {Swetz}, {Switzer},
  {Thornton}, {Visnjic}, \& {Wollack}}]{das/etal:2011}
{Das}, S., {Sherwin}, B.~D., {Aguirre}, P., {et~al.} 2011, \prl, 107, 021301,
  \dodoi{10.1103/PhysRevLett.107.021301}

\bibitem[{{Gualtieri} {et~al.}(2018){Gualtieri}, {Filippini}, {Ade}, {Amiri},
  {Benton}, {Bergman}, {Bihary}, {Bock}, {Bond}, {Bryan}, {Chiang}, {Contaldi},
  {Dor{\'e}}, {Duivenvoorden}, {Eriksen}, {Farhang}, {Fissel}, {Fraisse},
  {Freese}, {Galloway}, {Gambrel}, {Gandilo}, {Ganga}, {Gramillano},
  {Gudmundsson}, {Halpern}, {Hartley}, {Hasselfield}, {Hilton}, {Holmes},
  {Hristov}, {Huang}, {Irwin}, {Jones}, {Kuo}, {Kermish}, {Li}, {Mason},
  {Megerian}, {Moncelsi}, {Morford}, {Nagy}, {Netterfield}, {Nolta},
  {Osherson}, {Padilla}, {Racine}, {Rahlin}, {Reintsema}, {Ruhl}, {Runyan},
  {Ruud}, {Shariff}, {Soler}, {Song}, {Trangsrud}, {Tucker}, {Tucker},
  {Turner}, {List}, {Weber}, {Wehus}, {Wiebe}, \&
  {Young}}]{gualtieri/etal:2018}
{Gualtieri}, R., {Filippini}, J.~P., {Ade}, P.~A.~R., {et~al.} 2018, Journal of
  Low Temperature Physics, 193, 1112, \dodoi{10.1007/s10909-018-2078-x}

\bibitem[{{Hand} {et~al.}(2012){Hand}, {Addison}, {Aubourg}, {Battaglia},
  {Battistelli}, {Bizyaev}, {Bond}, {Brewington}, {Brinkmann}, {Brown}, {Das},
  {Dawson}, {Devlin}, {Dunkley}, {Dunner}, {Eisenstein}, {Fowler}, {Gralla},
  {Hajian}, {Halpern}, {Hilton}, {Hincks}, {Hlozek}, {Hughes}, {Infante},
  {Irwin}, {Kosowsky}, {Lin}, {Malanushenko}, {Malanushenko}, {Marriage},
  {Marsden}, {Menanteau}, {Moodley}, {Niemack}, {Nolta}, {Oravetz}, {Page},
  {Palanque-Delabrouille}, {Pan}, {Reese}, {Schlegel}, {Schneider}, {Sehgal},
  {Shelden}, {Sievers}, {Sif{\'o}n}, {Simmons}, {Snedden}, {Spergel}, {Staggs},
  {Swetz}, {Switzer}, {Trac}, {Weaver}, {Wollack}, {Yeche}, \&
  {Zunckel}}]{hand/etal:2012}
{Hand}, N., {Addison}, G.~E., {Aubourg}, E., {et~al.} 2012, \prl, 109, 041101,
  \dodoi{10.1103/PhysRevLett.109.041101}

\bibitem[{{Henderson} {et~al.}(2016){Henderson}, {Allison}, {Austermann},
  {Baildon}, {Battaglia}, {Beall}, {Becker}, {De Bernardis}, {Bond},
  {Calabrese}, {Choi}, {Coughlin}, {Crowley}, {Datta}, {Devlin}, {Duff},
  {Dunkley}, {D{\"u}nner}, {van Engelen}, {Gallardo}, {Grace}, {Hasselfield},
  {Hills}, {Hilton}, {Hincks}, {Hloẑek}, {Ho}, {Hubmayr}, {Huffenberger},
  {Hughes}, {Irwin}, {Koopman}, {Kosowsky}, {Li}, {McMahon}, {Munson}, {Nati},
  {Newburgh}, {Niemack}, {Niraula}, {Page}, {Pappas}, {Salatino}, {Schillaci},
  {Schmitt}, {Sehgal}, {Sherwin}, {Sievers}, {Simon}, {Spergel}, {Staggs},
  {Stevens}, {Thornton}, {Van Lanen}, {Vavagiakis}, {Ward}, \&
  {Wollack}}]{henderson/etal:2016}
{Henderson}, S.~W., {Allison}, R., {Austermann}, J., {et~al.} 2016, Journal of
  Low Temperature Physics, 184, 772, \dodoi{10.1007/s10909-016-1575-z}

\bibitem[{{Hlozek} {et~al.}(2019){Hlozek}, {Bond}, {Chapman}, {Chiang},
  {Dobbs}, {Fich}, {Foreman}, {Frolov}, {Halpern}, {Hinshaw}, {Murray},
  {Scott}, {Sievers}, \& {Vanderlinde}}]{hlozek/etal:2019}
{Hlozek}, R., {Bond}, J.~R., {Chapman}, S., {et~al.} 2019, in Canadian Long
  Range Plan for Astronony and Astrophysics White Papers, Vol. 2020, 50,
  \dodoi{10.5281/zenodo.3825611}

\bibitem[{{Lee} {et~al.}(2019){Lee}, {Abitbol}, {Adachi}, {Ade}, {Aguirre},
  {Ahmed}, {Aiola}, {Ali}, {Alonso}, {Alvarez}, \& et~al.}]{lee/etal:2019}
{Lee}, A., {Abitbol}, M.~H., {Adachi}, S., {et~al.} 2019, in Bulletin of the
  American Astronomical Society, Vol.~51, 147.
\newblock \doarXiv{1907.08284}

\bibitem[{{Mortillaro}(2020)}]{mortillaro:2020}
{Mortillaro}, N. 2020, CBC News.
\newblock
  \url{https://www.cbc.ca/news/technology/age-universe-science-cmb-1.5653450}

\bibitem[{{Racca} {et~al.}(2016){Racca}, {Laureijs}, {Stagnaro}, {Salvignol},
  {Lorenzo Alvarez}, {Saavedra Criado}, {Gaspar Venancio}, {Short}, {Strada},
  {B{\"o}nke}, {Colombo}, {Calvi}, {Maiorano}, {Piersanti}, {Prezelus},
  {Rosato}, {Pinel}, {Rozemeijer}, {Lesna}, {Musi}, {Sias}, {Anselmi},
  {Cazaubiel}, {Vaillon}, {Mellier}, {Amiaux}, {Berth{\'e}}, {Sauvage},
  {Azzollini}, {Cropper}, {Pottinger}, {Jahnke}, {Ealet}, {Maciaszek},
  {Pasian}, {Zacchei}, {Scaramella}, {Hoar}, {Kohley}, {Vavrek}, {Rudolph}, \&
  {Schmidt}}]{racca/etal:2016}
{Racca}, G.~D., {Laureijs}, R., {Stagnaro}, L., {et~al.} 2016, in Society of
  Photo-Optical Instrumentation Engineers (SPIE) Conference Series, Vol. 9904,
  Space Telescopes and Instrumentation 2016: Optical, Infrared, and Millimeter
  Wave, ed. H.~A. {MacEwen}, G.~G. {Fazio}, M.~{Lystrup}, N.~{Batalha},
  N.~{Siegler}, \& E.~C. {Tong}, 99040O, \dodoi{10.1117/12.2230762}

\bibitem[{{Sherwin} {et~al.}(2011){Sherwin}, {Dunkley}, {Das}, {Appel}, {Bond},
  {Carvalho}, {Devlin}, {D{\"u}nner}, {Essinger-Hileman}, {Fowler}, {Hajian},
  {Halpern}, {Hasselfield}, {Hincks}, {Hlozek}, {Hughes}, {Irwin}, {Klein},
  {Kosowsky}, {Marriage}, {Marsden}, {Moodley}, {Menanteau}, {Niemack},
  {Nolta}, {Page}, {Parker}, {Reese}, {Schmitt}, {Sehgal}, {Sievers},
  {Spergel}, {Staggs}, {Swetz}, {Switzer}, {Thornton}, {Visnjic}, \&
  {Wollack}}]{sherwin/etal:2011}
{Sherwin}, B.~D., {Dunkley}, J., {Das}, S., {et~al.} 2011, \prl, 107, 021302,
  \dodoi{10.1103/PhysRevLett.107.021302}

\bibitem[{{Swetz} {et~al.}(2011){Swetz}, {Ade}, {Amiri}, {Appel},
  {Battistelli}, {Burger}, {Chervenak}, {Devlin}, {Dicker}, {Doriese},
  {D{\"u}nner}, {Essinger-Hileman}, {Fisher}, {Fowler}, {Halpern},
  {Hasselfield}, {Hilton}, {Hincks}, {Irwin}, {Jarosik}, {Kaul}, {Klein},
  {Lau}, {Limon}, {Marriage}, {Marsden}, {Martocci}, {Mauskopf}, {Moseley},
  {Netterfield}, {Niemack}, {Nolta}, {Page}, {Parker}, {Staggs}, {Stryzak},
  {Switzer}, {Thornton}, {Tucker}, {Wollack}, \& {Zhao}}]{swetz/etal:2011}
{Swetz}, D.~S., {Ade}, P.~A.~R., {Amiri}, M., {et~al.} 2011, \apjs, 194, 41,
  \dodoi{10.1088/0067-0049/194/2/41}

\bibitem[{{Thornton} {et~al.}(2016){Thornton}, {Ade}, {Aiola}, {Angil{\`e}},
  {Amiri}, {Beall}, {Becker}, {Cho}, {Choi}, {Corlies}, {Coughlin}, {Datta},
  {Devlin}, {Dicker}, {D{\"u}nner}, {Fowler}, {Fox}, {Gallardo}, {Gao},
  {Grace}, {Halpern}, {Hasselfield}, {Henderson}, {Hilton}, {Hincks}, {Ho},
  {Hubmayr}, {Irwin}, {Klein}, {Koopman}, {Li}, {Louis}, {Lungu}, {Maurin},
  {McMahon}, {Munson}, {Naess}, {Nati}, {Newburgh}, {Nibarger}, {Niemack},
  {Niraula}, {Nolta}, {Page}, {Pappas}, {Schillaci}, {Schmitt}, {Sehgal},
  {Sievers}, {Simon}, {Staggs}, {Tucker}, {Uehara}, {van Lanen}, {Ward}, \&
  {Wollack}}]{thornton/etal:2016}
{Thornton}, R.~J., {Ade}, P.~A.~R., {Aiola}, S., {et~al.} 2016, \apjs, 227, 21,
  \dodoi{10.3847/1538-4365/227/2/21}

\end{thebibliography}
}
\end{multicols}

\end{document}